\documentclass[twocolumn,amsmath,amssymb]{revtex4}
\usepackage{graphicx}
\usepackage{dcolumn}
\usepackage{bm}

\newcommand{\etal}{{\it et al.}}
\begin{document}

\title{Surface Granular flows: Two Related Examples}

\author{D. V. Khakhar} \email{khakhar@iitb.ac.in}
\author{Ashish V. Orpe}
\affiliation{Department of Chemical
  Engineering, Indian Institute of Technology - Bombay, Powai, Mumbai,
    400076, India}
\author{J. M. Ottino} \affiliation{Department of Chemical and Mechanical Engineering, Northwestern University, Evanston, IL 60208, USA}

\begin{abstract}
Granular surface flows are common in industrial practice and natural
systems, however, theoretical description of such flows is at present
incomplete. Two prototype systems involving surface flow are compared:
heap formation by pouring at a point and  rotating cylinders. Continuum
models for analysis of these flows are reviewed, and experimental
results for quasi-2d systems are presented. Experimental results in both
systems are well described by continuum models.
\end{abstract}

\maketitle

\section{Introduction} \label{intro}
\begin{figure*}
\centering{\includegraphics[width=5in]{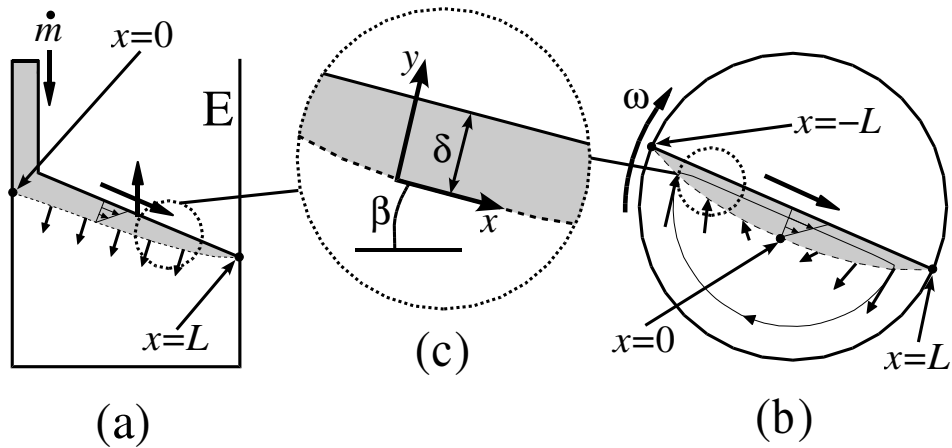}}
\caption{Schematic view of surface flow systems: (a) Heap flow (b)
Rotating cylinder flow.  (c) Coordinate system used in the analysis.
\label{schematic}}
\end{figure*}

Granular flows have been the subject of considerable recent work [1-5] driven
by both technological needs \cite{enn94,bri95} and the recognition that many
aspects of the basic physics are poorly understood \cite{deg99}. Surface flows
of granular materials, that is flows confined to a surface layer on
a static granular bed, are important in industrial practice and nature.
Industrial examples appear in the transportation, processing and storage
of materials in systems such as rotary kilns, tumbling mixers, and
feeding and discharge of silos. Examples in nature include the formation
of sand dunes, lava flow, avalanches, and transport of sediments in
rivers.  Although considerable progress has been made, theoretical
description of surface flows is incomplete at present. Several
approaches, based on different assumptions about the physics of the
flows, have been proposed [9-19].
A few experimental studies are also available [9,19-33].
Most work is focussed on two systems: heap flow and rotating
cylinder flow shown schematically in Fig.~\ref{schematic}. 

An important feature of surface granular flows is the interchange of
particles between the flowing layer and the fixed bed. In the case of a
rotating cylinder  the interchange rate is determined by kinematics
since the velocity of the fixed bed at the bed-layer interface is known.
The situation in the case of heap flow is more complicated. Bouchaud
{\it et al.} \cite{bouc94} proposed a phenomenological model (BCRE
model) in which the interchange rate is determined by the local surface
angle. A variation of this model proposed by Boutreux {\it et al.}\
\cite{bout98} (BRdG model) has been broadly validated by continuum
models \cite{dou99,kha01} and experiments \cite{kha01}, as we show
below. Continuum models developed previously, for both heaps and
rotating cylinders, are all based on depth-averaged hydrodynamic
equations and differ primarily in the constitutive equations used. All
the models contain parameters which must be evaluated from experiments,
but in most cases, these parameters have not been determined.

Here we present here a common continuum based framework for the analysis of 
both heap flow and rotating cylinder flow. The treatment closely follows that
given in refs.~\cite{kha01} and \cite{orp01}. Model predictions are
compared to experimental results and to predictions of previous models. 
The general continuum model is presented first. Results
for the heap formation and rotating cylinder flow are given next
followed by conclusions.

\section{General continuum model}\label{conti}

Consider a flowing layer on the surface of a granular bed assuming the flow is
nearly uni-directional in the layer and curvature effects are small. The depth
averaged continuity equation
and the $x$-momentum balance equation are simplified using the following 
assumptions. The bulk density in the layer ($\rho$) is 
nearly constant (since the dilation of the flowing particles is not too
large in the relatively slow flows being considered).
The velocity
profile in the layer is linear and of the form [31,32] $v_x=2u(y/\delta)$,
where $u(x,t)$ is the depth averaged velocity in the
layer and $\delta$ is the layer thickness. Slow plastic deformation [33] is 
neglected. 
The shear stress at the interface is taken to be \cite{orp01}
\begin{equation} \label{stress}
\tau_{xy}|_{y=0}=\rho d^{2} f(\rho) \left(\frac{\partial v_x}{\partial y}\right)^2
-\rho g \delta\cos\beta\tan\beta_s
\end{equation} where $d$ is the
particle diameter and $\tan\beta_s$ is  the effective coefficient of {\it
dynamic} friction, with $\beta_s$ taken to be the static angle of
repose. The stress is sensitively dependent on the local bulk density and 
based on recent empirical evidence \cite{orp01} we take $f(\rho) = c\delta/d$ 
with $c\approx1.5$.
The governing equations then reduce to
\begin{eqnarray} \label{contf} \frac{\partial\delta}{\partial
t}+\frac{\partial\ }{\partial x}\left(\delta u\right)&=&-\Gamma,\\ \label{momf}
\frac{\partial\ }{\partial t}\left(\delta u\right) +
\frac{4}{3}\frac{\partial\ }{\partial x}\left(\delta u^2\right)&=&
-4cd\frac{u^2}{\delta}+ g\delta\frac{\sin(\beta-\beta_s)}{\cos\beta_s},
\end{eqnarray}
where $\Gamma$ is the flux from the layer into the bed.
Further, assuming the static friction forces at the heap-layer interface
to be fully mobilized, the Mohr-Coulomb criterion yields
\begin{equation}\label{mc}
\tau_{xy}|_{y=0}=-\rho g\delta\cos\beta\tan\beta_m,
\end{equation}
where $\tan\beta_m$ is the effective coefficient of {\it static}
friction. 
Using eq.~(\ref{stress}) and the assumptions given above, eq.~(\ref{mc}) yields
\begin{equation}\label{mc3}
u=\dot{\gamma}\delta/2, \end{equation}
with the shear rate given by
\begin{equation}\label{gamdot}
\dot{\gamma}=\left[\frac{g\cos\beta\sin(\beta_m-\beta_s)}
{cd\cos\beta_m\cos\beta_s}\right]^{1/2}.
\end{equation}
A similar analysis is given by Douady {\it et al.}\ \cite{dou99} with the 
difference that no stress constitutive equation such as eq.~(\ref{stress}) is 
used and instead the shear 
rate in the flowing layer is assumed to be constant.

\section{Heap formation}\label{heapf}
Consider a quasi-steady flow ($\partial \delta/\partial t,\partial u/\partial
t\approx 0$) and a slowly varying interface angle ($\partial
\beta/\partial x\approx 0$) during heap formation. 
The continuity equation (eq.~\ref{contf}) together with eq.~(\ref{mc3}) then 
reduces to 
\begin{equation} \label{qsscont}
\dot{\gamma}\delta\frac{\partial \delta}{\partial x}=-\Gamma,
\end {equation}
and the momentum balance equation (eq.~\ref{momf}) together with
eq.~(\ref{mc}) simplifies to
\begin{equation}  \label{qssmom}
\dot{\gamma}^2\delta\frac{\partial \delta}{\partial x}=
-\frac{g\sin(\beta_m-\beta)}{\cos\beta_m} .
\end{equation}
Combining eqs.~(\ref{qsscont}) and (\ref{qssmom}) yields
$\Gamma=g\sin(\beta_m-\beta)/\dot{\gamma}\cos\beta_m$,
which, for the case when $\beta_m\approx\beta$, reduces to
\begin{equation}   \label{source2}
\Gamma\approx V(\beta_m-\beta),
\end{equation}
where $V=g/\dot{\gamma}\cos\beta_m$. Thus, the continuum model
yields a source term similar to the BRdG model; the scaling of $V$ is
also similar to the BRdG model.

\begin{figure*}
\centering{\includegraphics[width=6.5in]{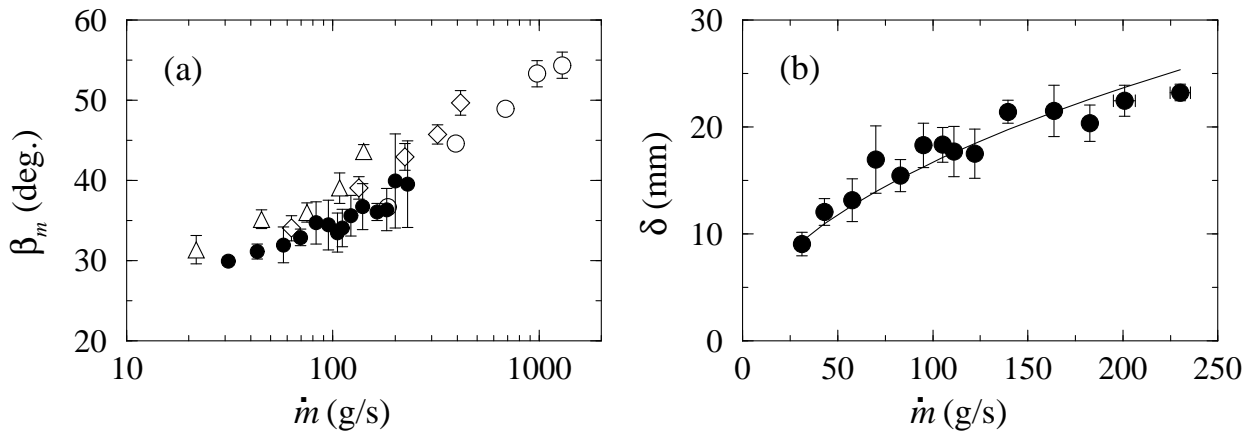}}
\caption{Variation of the (a) maximum angle of repose ($\beta_m$) and (b) 
layer thickness ($\delta$) with mass
flow rate ($\dot{m}$) for 2 mm steel balls. Filled symbols: open heap
system [19]. Open symbols: rotating cylinder system for three
cylinder sizes [30]. The solid line in (b) is a best fit of the form 
$\delta\propto \dot{m}^{1/2}$ \label{betamdelta.mdot}}
\end{figure*}

We further simplify the above equations for two different geometries of
heap formation: {\it closed}, as shown in Fig.~\ref{schematic}a, and
{\it open} in which the end wall (E, Fig.~\ref{schematic}a) is removed.
In the open system at steady state, all the
material entering the system leaves at the far edge of the heap and no
particles are absorbed or eroded. This implies that $\Gamma=0$, which on
substituting into eq.~(\ref{source2}) $\beta=\beta_m\equiv$ constant.
Using these results in eqs.~(\ref{qsscont}) and (\ref{qssmom})
shows that the average velocity ($u$) and thickness ($\delta$) of the
flowing layer are also constant in open systems. The mass flow rate in
the system is given by $\dot{m}=\rho u \delta T$, where $T$ is the width
of the layer. This expression, together with eq.~(\ref{mc3}), gives the
following relationship between the layer thickness and mass flow rate
\begin{equation}\label{opendelta}
\delta=\left[2\dot{m}/(T\rho\dot{\gamma})\right]^{1/2}.
\end{equation}
Experimental results \cite{kha01} based on flow
visualization studies validate the above predictions, and sample
results are given below. Fig.~\ref{betamdelta.mdot}a shows the variation of
the maximum angle of repose with mass flow rate in the system for 2 mm
steel balls in an open heap system (filled symbols).  The data indicate
that $\beta_m$, and thus the coefficient of static friction at the
heap-layer interface ($\tan\beta_m$), is not a constant but increases
with the local flow rate. An increase in surface angle with flow rate
was also reported by Lemieux and Durian \cite{lem00}. 
Fig.~\ref{betamdelta.mdot}b shows the variation of the layer thickness
($\delta$) with mass flow rate. The solid line is a fitted curve of the
form $\delta\propto\dot{m}^{1/2}$. This indicates agreement with
theoretical predictions (eq.~\ref{opendelta}) if the product
$\rho\dot{\gamma}$ is independent of mass flow rate.

In a closed system (Fig.~\ref{schematic}a), at steady
state we must have $\Gamma\equiv$ constant for the heap to rise
uniformly. Integrating eq.~(\ref{qsscont}), the layer thickness profile
is obtained as
\begin{equation}  \label{closeddelta}
\delta=\left[\delta_L^2+2\Gamma(L-x)/\dot{\gamma}\right]^{1/2}
\end{equation}
where $\delta_L$ is the layer thickness at the end of the layer, $x=L$,
and $L$ is the length of the interface (Fig.~\ref{schematic}a). The
rise velocity is related to the mass flow rate by
\begin{equation} \label{closedGamma}
\Gamma=\dot{m}/(TL\rho),
\end{equation}
and the interface angle is calculated from eq.~(\ref{source2}).

\begin{figure*}
\centering{\includegraphics[width=6.5in]{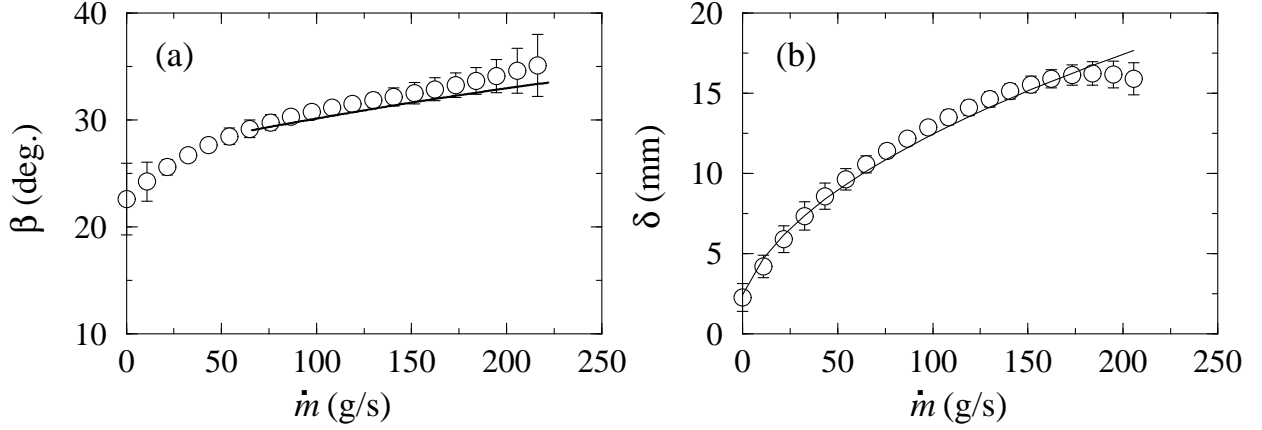}}
\caption{Variation of the (a) surface angle ($\beta$) and (b) layer thickness
($\beta$) with distance from the edge of the heap ($L-x$) for
flow of $2$ mm steel balls in a closed system. Symbols are experimental data 
and error bars indicate the standard deviation over six measurements. Solid 
line in (a) is a fit of eq.~(\ref{source2}) and in (b) is the prediction of
eq.~(\ref{closeddelta}).\label{deltabeta.x}}
\end{figure*}

Experimental results \cite{kha01} for
closed systems show that the rise velocity ($\Gamma$)  varies nearly
linearly with mass flow rate in agreement with eq.~(\ref{closedGamma}),
and the bulk density, which is found to be constant, is $\rho=3.2$ g/cm$^3$.
Fig.~\ref{deltabeta.x} shows the variation of both interface angle ($\beta$)
and layer thickness ($\delta$) with length along the bed-layer
interface ($L-x$) for a fixed mass flow rate. The solid line in
Fig.~\ref{deltabeta.x}b is a fit of eq.~(\ref{closeddelta}).
There is a good match between the
fitted line and the experimental data, which suggests that the shear
rate, $\dot{\gamma}$, is constant. Similar results are obtained for
all flow rates studied. Using experimental results for the rise velocity
($\Gamma$) and the interface length ($L$), we obtain $\dot{\gamma}=20\pm
2$ s$^{-1}$ from eq.~(\ref{closeddelta}), where the standard deviation
indicated is calculated for all 10 flow rates studied.  Using the value
of the bulk density obtained above, we find from eq.~(\ref{opendelta})
that the shear rate for the open system is $\dot{\gamma}=22\pm 3$
s$^{-1}$. The value of the shear rate predicted from eq.~(\ref{gamdot}) is
$\dot{\gamma}=20\pm 5$ s$^{-1}$ for the range of mass flow rates
considered. Thus the shear rates for the open and closed systems are the
same within experimental error, and predictions of theory are in
reasonable agreement with experimental values.

\section{Rotating cylinder}\label{rotat}
The simplest case corresponds to rotating cylinder flow for 50\% fill 
fraction. Assuming a nearly flat interface, the source term is given by
$\Gamma=\omega x$. Substituting into the continuity
equation (eq.~\ref{contf}) and integrating we obtain
\begin{equation} \label{udelta}
u\delta=\frac{\omega}{2}\left(L^2-x^2\right).
\end{equation}
Using eq.~(\ref{contf}) the momentum balance equation
(eq.~\ref{momf}) simplifies to
\begin{equation}  \label{ududx}
u\frac{du}{dx}=\frac{3g}{4}\frac{\sin(\beta-\beta_s)}{\cos\beta_s}
-3\frac{cdu^2}{\delta^2}+ \frac{\omega x u}{\delta}.
\end{equation}
We consider two different limiting solutions to eqs.~(\ref{udelta}) and
(\ref{ududx}) below.

Firstly, consider the case when shear rate ($\dot{\gamma}$) is nearly
constant. Using eq.~(\ref{mc3}), the flux equation (eq.~\ref{udelta})
gives the layer thickness profile as
\begin{equation}\label{deltamak}
\delta=\left(\frac{\omega}{\dot{\gamma}}\right)^{1/2}
\left(L^2-x^2\right) ^{1/2},
\end{equation}
which is symmetric for all rotational speeds ($\omega$).
Eq.~(\ref{deltamak}) corresponds to the model of Makse \cite{mak99}, in
which the shear rate is assumed to be a fitting parameter. In the
present case the shear rate is obtained from eq.~(\ref{gamdot}) and the 
mean velocity is given by $u=\dot{\gamma}\delta/2$.
Substituting these results in eq.~(\ref{ududx}), and using the
Mohr-Coulomb condition (eq.~\ref{mc}) yields eq.~(\ref{source2}) with
$\Gamma=\omega x$ and $\beta_{m}\approx\beta$. This allows for calculation of 
the angle ($\beta$) along the interface. Thus the assumption of a constant 
shear rate is consistent with the model, and gives a complete description of
the flow. However, it is not apparent from the analysis, under what
conditions the solution is valid.

Consider next the case when the shear rate is not constant along the
layer, but when the acceleration ($du/dx$) is small. Eliminating $\delta$ using
eq.~(\ref{udelta}), the scaled momentum balance becomes
\begin{equation}\label{momscal}
\bar{u}\frac{d\bar{u}}{d\xi}=
\frac{3}{4Fr}\frac{\sin(\beta-\beta_s)}{\cos\beta_s}
-12cs\frac{\bar{u}^4}{(1-\xi^2)^2}+\frac{2\xi\bar{u}^2}{1-\xi^2},
\end{equation}
where $\bar{u}=u/\omega L$, $\xi=x/L$ and the dimensionless parameters are the
Froude number, $Fr=\omega^2L/g$, and the size ratio, $s=d/L$. The first term on
the right hand side of eq.~(\ref{momscal}) is the net driving force, that is 
the gravitational force less the frictional resistance to flow, and is
independent of the flow velocity ($\bar{u}$). The second term is the `viscous'
resistance due to collisional stresses, and the third term arises as a result 
of in-flow and out-flow of particles from the layer. Both these terms depend 
on the flow velocity. Typical experimental Froude numbers for experiments in 
rotating
cylinders are in the range $O(10^{-3})$ to $O(10^{-2})$. In these cases the
driving force term ($O(1/Fr)$) is much larger than the acceleration term
($O(\xi/\sqrt{sFr})$ based on eq.~\ref{mc3}), particularly near the midpoint of
the layer ($\xi=0$). The collisional stress term is of the same magnitude as 
the
net driving force term since the flow velocity increases to balance the two.
Thus for $\xi\sqrt{Fr/s}\ll 1$ the acceleration term may be neglected.

For negligible acceleration ($d\bar{u}/d\xi\approx 0$), the scaled mean flow
velocity is  obtained from eq.~(\ref{momscal}) as
\begin{equation}\label{meanvel}
\bar{u}=\left(\frac{1-\xi^2}{12cs}\right)^{1/2}
\left[\xi+(\xi^2+9csA/Fr)^{1/2}\right]^{1/2},
\end{equation}
where $A=\sin(\beta-\beta_s)/\cos\beta_s$. Using eq.~(\ref{udelta}), the 
scaled layer thickness profile is
\begin{equation}\label{deltath}
\bar{\delta}=\left[\frac{3cs(1-\xi^2)}{\xi+(\xi^2+9csA/Fr)^{1/2}}\right]^{1/2},
\end{equation}
where $\bar{\delta}=\delta/L$. The above solution is valid only if $A>0$, that
is if $\beta>\beta_s$. For $\beta\le\beta_s$, we have $\bar{u}=\bar{\delta}=0$,
thus there is no steady flow possible if the interface angle is less the static
angle of repose. This is consistent with the definition of the static angle of
repose. Note that the layer profile is not symmetric about $\xi=0$, and for any
$\xi>0$ we have $\bar{\delta}(-\xi)>\bar{\delta}(\xi)$, that is, the upper part
of the layer ($\xi<0$) is thicker than the lower part. The source of the
asymmetry is the in-flow/out-flow term in the momentum balance
(third term on the right hand side of eq.~\ref{momscal}). In the upper
part of the layer ($\xi<0$) the flow is retarded by material entering
the layer from the bed ($\Gamma<0$) and the reverse is true in the lower
part of the layer. Thus, the layer is thicker in upper part because of
the lower velocity relative to the lower part of the layer ($\xi>0$), resulting
in a skewed profile. Further, eq.~(\ref{deltath}) indicates that the profile
becomes more skewed with increasing Froude number ($Fr$) and decreasing
size ratio ($s$). In the limit, $Fr/s\ll 1$, the scaled layer
thickness profile becomes $\bar{\delta}=(csFr/A)^{1/4}(1-\xi^2)^{1/2}$,
which is identical to the result obtained assuming a constant
shear (eq.~\ref{deltamak}) when eq.~(\ref{gamdot}) is used to calculate
the shear rate. This implies that a profile symmetric about the layer
midpoint ($\xi=0$) is obtained at very low Froude numbers and relatively high 
size ratios, and in this
limit the shear rate is nearly constant.

\begin{figure*}
\centering{\includegraphics[width=6.5in]{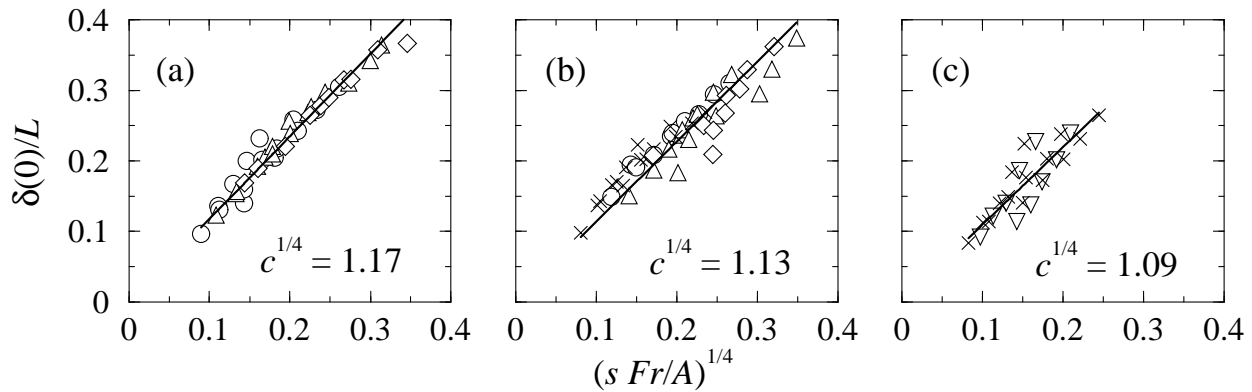}}
\caption{Variation of the layer thickness at the midpoint ($\delta(0)$)
with $(sFr/A)^{1/4}$ for (a) steel balls, (b) glass beads (c) and sand 
particles in cylinders of different sizes and at different
rotational speeds. Symbols are experimental data for different sized particles:
{\large$\circ$} $d = 1$ mm, {\scriptsize$\triangle$} $d = 2$ mm,
{\large$\diamond$} $d = 4$ mm, {\scriptsize$\nabla$} $d = 0.4$ mm and
{\large$\times$} $d = 0.8$ mm.
The solid line is a fit of eq.~(\ref{deltath}) and the values of the 
parameter $c^{1/4}$ are indicated.} \label{delta.sFr} \end{figure*}

The interface angle profile ($\beta(\xi)$) is obtained 
from eq.~(\ref{source2}), using $\Gamma=\omega x$ and
eq.~(\ref{gamdot}) as
\begin{equation} \label{betath}
\beta(\xi)=\beta_m- \frac{Fr \cos{\beta_m}}{3 c s} \left[\xi + 
(\xi^2 + 9 c s A /Fr)^{1/2}\right]^{1/2} \xi.
\end{equation}
In simplifying the preceding equation we assume
$\beta_m\approx\beta\approx\beta_s$. Eq.~(\ref{betath})
indicates that the interface angle decreases monotonically with
distance along the interface and at $\xi=0$, $\beta=\beta_m$. Thus in
the rotating cylinder flow the maximum angle of repose can be
experimentally obtained by measuring the interface angle at the midpoint
of the layer. For $(Fr/s)\xi$ sufficiently large and $\xi>0$, we
get $\beta<0$, that is, for small size ratios and large Froude numbers
the layer profile may turn up at the end. Conversely, when
$(Fr/s)\ll 1$, eq.~(\ref{betath}) yields $\beta\approx\beta_m$,
and the interface profile is nearly flat. Neglecting terms
$O(\xi Fr/s)$, which is consistent with the approximation in the
momentum balance equation, we get $A=(\beta_m-\beta_s)/\cos\beta_s$.

\begin{figure*}
\centering{\includegraphics[width=6.5in]{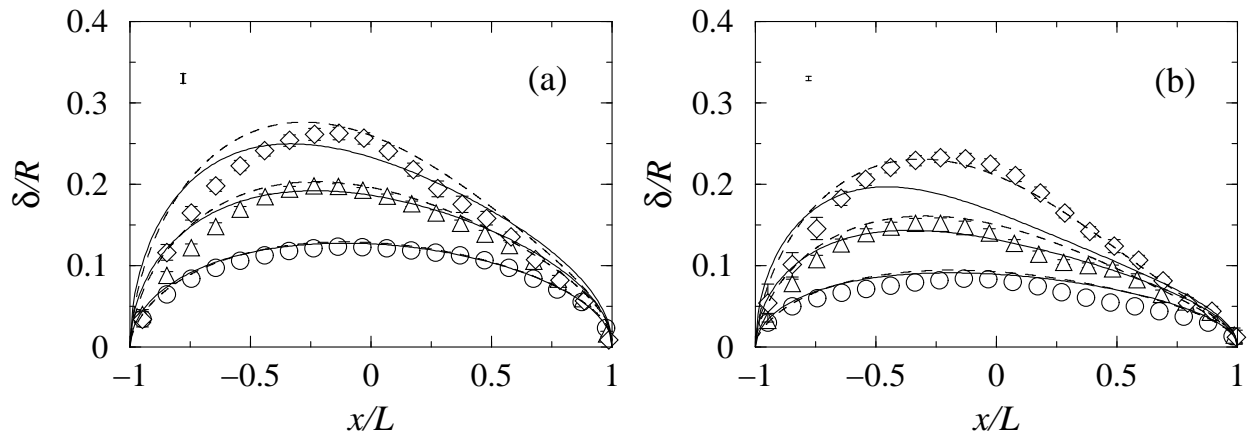}}
\caption{Layer thickness profiles for (a) $2$ mm steel balls and (b) $0.8$
mm sand. Symbols denote experimental data for three different
froude numbers ($Fr$): {\large$\circ$}
$Fr = 2 \times 10^{-3}$, {\scriptsize$\triangle$} $Fr = 22 \times 10^{-3}$,
{\large$\diamond$} $Fr = 64 \times 10^{-3}$. Solid lines are predictions
of eq.~(\ref{deltath}) and dashed lines are the predictions of the model of 
Khakhar \etal\ [13]. The error bars give the standard deviation over
6 measurements and the bar indicates the scaled diameter of a particle
($s=d/R$). \label{delta.xi}}
\end{figure*}

Consider next a comparison of the theoretical results to experimental data.
A few key numbers are reported, as they convey a sense of qualitative 
agreement. However, for full details the reader is referred to \cite{orp01}. 
The model parameters required are $\beta_s$, $\beta_m$  and $c$.
Data of Orpe and Khakhar \cite{orp01} for the
first two parameters are shown in Fig.~\ref{betamdelta.mdot}a (open symbols) 
for 2 mm steel
balls in rotating cylinders of 3 sizes and for different rotational
speeds of the cylinders. The data correlates reasonably well with the mass flow
rate at the midpoint of the layer calculated from $\dot{m}=\rho\omega
L^2T/2$, where $T$ is the cylinder length and the same density as in the
heap experiments ($\rho=3.2$ g/cm$^3$) is used. Data
spanning nearly two decades of flow rate fall on a single curve,
although with some scatter. The maximum angle of
repose increases with mass flow rate, and the measured values are similar to
those from heap experiments which are also shown in the same figure. The static
angle of repose is the angle at $\dot{m}=0$.

Orpe and Khakhar \cite{orp01} had obtained $c\approx 1.5$ by fitting the
theory of Khakhar \etal\ \cite{kha97} to experimental layer thickness
profiles. We obtain a new estimate of the parameter  based on the
layer thickness at the midpoint of the layer ($\xi=0$), which, from
eq.~(\ref{deltath}), is $\bar{\delta}(0)=(csFr/A)^{1/4}$. Fig.~\ref{delta.sFr}
shows experimental data for $\bar{\delta}(0)$ versus $(sFr/A)^{1/4}$ for
experimental data for 90 experiments comprising steel balls, glass beads
and sand of different sizes in cylinders of different sizes and for
different rotational speeds. The data falls on a straight line for each
material (although with some scatter) and a least squares fit gives $c=1.9$
for steel balls, $c=1.6$ for glass beads and $c=1.4$ for sand. Since the model
is essentially exact at $\xi=0$, the good fit implies that the proposed
constitutive equation for stress is reasonable, and the shear rate in the layer
is well-described by eq.~(\ref{gamdot}) at $\xi=0$. 

Predictions of the model for the layer thickness profile are compared to
experimental data in Fig.~\ref{delta.xi} for sand particles and steel
balls for different Froude numbers in a cylinder of radius 16 cm, using
the value of $c$ obtained above and experimental values for $\beta_m$
and $\beta_s$. The agreement is good except at the highest $Fr$ and low $s$, and
the all the qualitative features of the data are reproduced. At low $Fr$ and
relatively high $s$ studied, the profile is nearly symmetric (steel balls at 
the lowest
$Fr$), and the profiles become more skewed with increasing $Fr$ and decreasing
$s$. The deviation at the high Froude numbers and low size ratio are due to
neglect of the acceleration term. Similar agreement is obtained for the other
cases studied as well. The predictions of the model of Khakhar \etal\
\cite{kha97} are shown in the figure as dashed lines. These nearly coincide with
the results from the present model, except for the highest $Fr$ for
sand, indicating that the approximations made are reasonable for the parameter
values of interest. It is remarkable that such a simple theory is able to
describe the behavior of the system over such a wide range of parameters:
materials include steel balls, glass beads and sand; varying shapes with steel
balls being spherical, glass beads, nearly spherical and sand being irregularly
shaped; size ratios in the range $s\in(0.005,0.05)$ and Froude numbers in the
range $Fr\in(2\times10^{-3},64\times10^{-3})$. Model predictions of the 
interface angle profile are in reasonable agreement with experiments 
\cite{orp01}.

\section{Conclusions}\label{concl}
A theoretical framework serves to unify the behaviour of surface flows
for two prototypical systems: heap flow and
rotating cylinder flow. The model is based on a stress constitutive
equation and failure criterion which contain three material parameters:
$\beta_s$, $\beta_m$ and  $c$.  Analytical results for both systems give 
a complete description 
of the systems  in terms of the layer thickness profiles ($\delta(x)$),
average velocity of flow ($u(x)$)  and the interface angle profile
($\beta(x)$). In open heap systems a  layer of uniform thickness with a
uniform flow velocity is obtained, whereas in the closed heap system
$\delta^2\propto x$.  The interface angle is constant and equal to the
maximum angle of repose in the open system, whereas it decreases with
distance from the pouring point in the closed system. Results for the
rotating cylinder are obtained for the case when the acceleration of
particles in the layer is small ($\xi\sqrt{Fr/s}\ll 1$).  The layer
profile is found to be asymmetric about the midpoint of the layer
($\xi=0$) with the upper part of the layer ($\xi<0$) being thicker. The
skewness  increases with increasing Froude numbers and decreasing size
ratios. The scaled shear rate ($\dot{\gamma}/\omega$) decreases with
increasing Froude number and size ratio. 
The layer interface angle decreases with distance in
the flow direction. For high $\xi Fr/s$ and $\xi>0$ the layer turns up, whereas
when $\xi Fr/s$ is small a nearly flat interface is obtained.

Quasi-2d experiments carried out for open and closed heaps and rotating
cylinders of different sizes, by and large, validate the predictions of the
theory. The three material parameters of the model ($\beta_{s}$, $\beta_{m}$ 
and $c$) are all obtained from relatively simple measurements. The model 
equation can thus be applied to more complex geometries. Deviations of the 
model from experimental data appear in the 
interface angle profile in the rotating cylinder flow. 
This is most likely due to end wall effects which are discussed in 
\cite{orp01}.

\section*{Acknowledgements}
D.~V.~Khakhar acknowledges the financial support of the Department of Science
and Technology, India, through the Swarnajayanti Fellowship project
(DST/SF/8/98) for part of this work. This work was supported in part by grants
to J. M. Ottino from the Division of Basic Energy Sciences of the Department of
Energy, the National Science Foundation, Division of Fluid and Particulate
Systems, and the Donors of the Petroleum Research Fund, administered by the
American Chemical Society.

\end{document}